\documentclass[12pt]{article}
\usepackage{epsf}
\newcommand{\eqb}{\begin{equation}}
\newcommand{\eqe}{\end{equation}}
\newcommand{\dmb}{\begin{displaymath}}
\newcommand{\dme}{\end{displaymath}}
\newcommand{\pd}{\partial}
\newcommand{\ep}{\varepsilon}
\newcommand{\eab}{\begin{eqnarray}}
\newcommand{\eae}{\end{eqnarray}}
\newcommand{\ra}{\right\rangle}
\newcommand{\la}{\left\langle}
\newcommand{\e}{\mbox{e}}
\newcommand{\be}{\begin{equation}}
\newcommand{\ee}{\end{equation}}

\def\beq{\begin{equation}}
\def\eeq{\end{equation}}
\def\pl{\partial}

\def\al{\alpha}
\def\bt{\beta}

\def\de{\delta}

\def\si{\sigma}
\def\Si{\Sigma}
\def\te{\theta}
\def\Te{\Theta}
\def\La{\Lambda}
\def\lam{\lambda}

\def\ep{\epsilon}
\def\vp{\varphi} 
\def\sq{\sqrt}

\def\l{\left (}
\def\rq{\right ]}
\def\lq{\left [}
\def\r{\right )}
\def\fr{\frac}
\def\hs{\hspace}

\def\inf{\infty}
\def\ran{\rangle}
\def\lan{\langle}
\def\ov{\overline}

\begin{document}
\begin{titlepage}
\begin{flushright}
HD-THEP-03-24\\
May 21, 2003
\end{flushright}
\vspace{0.6cm}
\begin{center}
{\Large \bf Supersymmetric Models For Gauge Inflation} 
\end{center}
\vspace{0.5cm}

\begin{center}
{\large   
R. Hofmann$^{\diamond }$,
F. Paccetti Correia$^{\diamond }$,
M. G. Schmidt$^{\diamond }$,\\
\vspace{0.1cm}
Z. Tavartkiladze$^{\diamond ,1}$ 
}

\vspace{0.3cm}

{\em 
Institut f\"ur Theoretische Physik,
Universit\"at Heidelberg\\ 
Philosophenweg 16, 69120 Heidelberg, Germany}
\end{center}
\vspace{0.4cm}
\begin{abstract}

We present possible realizations of gauge inflation arising from a 5D
${\cal N}=1$ supersymmetric $U(1)$ model, where the extra dimension is
compactified on a circle. A one-loop inflaton effective 4D potential is
generated, with the inflaton being a
'Wilson-line field'. It relies on a SUSY breaking. We first consider 
SUSY breaking to occur spontaneously within a 'no-scale' model by a non zero $F$-term 
of the radion
superfield which transmits SUSY breaking into the 'visible' sector. 
As an alternative, we study
$D$-term SUSY breaking originating directly from the 5D gauge
supermultiplet. 
Together with the usual KK resummation method, we present a calculation 
with the
world-line formalism. The latter allows one to get the resulting effective
potential directly as a sum over all winding modes.
For both presented scenarios, the generated effective
potentials have suitable forms for realizing successful inflation,
i.e. are flat enough and give the needed number of ${\rm e}$-foldings. In
addition, there is a natural way to get strongly suppressed values for
the potentials, which then could be associated with 
dark energy/quintessence.

\end{abstract} 
\vspace{2cm}
\footnoterule

{\small
\noindent$^\diamond $E-mail addresses: R.Hofmann@, F.Paccetti@, M.G.Schmidt@,

\hspace{2.6cm}Z.Tavartkiladze@ThPhys.Uni-Heidelberg.DE\\
$^1$On leave of
absence from Institute of Physics, GAS, Tbilisi 380077, Georgia.} 

%

\end{titlepage}

\section{Introduction}

Inflation is an efficient way to solve the 
cosmological flatness, horizon, and monopole 
problem \cite{Guth}. It explains naturally why the 
visible part of the universe appears to be 
isotropic on large scales. Field theoretic real-time models of 
inflation can explain the seeding of structure formation. 
The experimental data on the spectral 
properties of primordial density perturbations do not lead to a unique inflation 
model in field theory. If in a given field theory model the scale 
of inflation is close to the 4D Planck 
scale $M_P$ the slow-roll requirements linked to 
the model's tree-level potential generically are violated by the appearance of 
gravitationally induced operators. Even for low-scale 
inflation the coupling of the inflaton to other fields induces 
radiative corrections to the inflaton potential 
which are hard to control. This problem is much milder if 
a symmetry protects the potential. For example spontaneously broken, rigid
supersymmetry (SUSY) 
yields flat potentials \cite{copel} for inflaton fields if the scale 
of SUSY breaking is much lower than $M_P$. Another possibility for the generation of 
radiatively stable, flat potentials is the spontaneous breakdown 
of a global symmetry at scale $f$. Very recently, this possibility was 
used to construct Little Higgs theories 
\cite{LittleHiggs}. Furthermore, spontaneously broken global symmetries may underly 
the curvaton scenario for the non adiabatic generation 
of curvature perturbations after inflation \cite{curvaton}. 
Coming back to inflation itself, it is well known
that a small amount of explicit symmetry breaking on top of the spontaneous breakdown 
opens up the possibility that (quasi) de-Sitter-cosmology can 
be driven by the associated pseudo Nambu-Goldstone (PNG) fields \cite{Freese}. 
Slow roll of these fields is ensured if $M_P/f\ll 1$. 
However, a symmetry breaking above $M_P$ is probably 
beyond a field theoretical treatment. This significantly reduces the 
appeal of the 4D PNG model. 
This problem was recently pointed out in \cite{AHRandall}, 
and a resolution in terms of gauge symmetry combined 
with the assumption of extra, compactified dimensions was proposed. 
In this context the possible generation of a varying spectral index of 
the primordial, adiabatic scalar perturbations was investigated 
in ref.\,\cite{Zhang}. WMAP data \cite{WMAP} indicate a tilted spectrum.

A 4D effective potential for the Wilson-line 
phase 
\eqb
\Te\equiv \int_0^{2\pi R} dx^5 A_5\,,
\eqe
($R$ denotes the compactification radius) 
arises by means of the Hosotani mechanism \cite{Hosotani} if 
charged states are present in a theory. The potential is protected from local 
quantum gravity corrections due to 5D gauge invariance. Moreover, the scale $f_\Theta$, 
which enters the slow-roll condition $M_P/f_\Theta\ll 1$, is naturally larger 
than $M_P$ if the effective 4D gauge theory is weakly coupled. 
A variant of this model uses a mass $M\gg R^{-1}$ for 
the charged bulk fluctuations which exponentially suppresses 
the value of the potential \cite{AHRandall}  making it a suitable 
candidate \cite{Riotto} for dark energy i.e. quintessence \cite{quint}. 
To derive the effective 4D potential for $\Theta$ one usually appeals 
to `Kaluza-Klein' (KK) regularization, i.e. a Poisson 
resummation of the KK spectrum \cite{Barb}. 
This recipe avoids the use of a cutoff which would destroy some of 
the symmetries of 
the underlying higher dimensional theory. 
Recently, KK regularization was intensively disputed, 
and it was pointed out that SUSY and gauge symmetry in 
field theory help to make KK regularization consistent but that ultimately it is 
string theory that provides a physical KK completion in the ultraviolet \cite{Nilles}. 
It may be helpful in this context to point out that in a world-line 
formulation of quantum field theory \cite{WLF} in 5D spacetime no 
Poisson resummation of the KK tower is needed, 
since directly a sum over winding modes is introduced in the action \cite{therm}
$$
-\int d^4xdx_5\int_0^\infty
\frac{dT}{T}\sum_{k=-\infty}^{\infty}
\int {\cal D}^5y
\exp\left[-\int_0^T d\tau \left(\frac{(\dot{x}_M)^2}{4}+iq\dot{x}_5
A_5+M^2\right)\right]\,,
$$
\beq
{\rm with}~~x_M(\tau)=2\pi R\,k\frac{\tau}{T}\delta_{M5}+y_M(\tau)+x_M\,,\
\ \ 
\int_0^Td\tau y_M(\tau )=0~,
\label{WLTHintr} 
\eeq
(see Appendix B for details). 
In the non-SUSY case this approach exhibits a 5D UV divergence from 
the zero-winding 
sector in the one-loop effective potential which is cut off by restricting 
the proper-time integration to $T\ge \Lambda^{-2}$.  

In a SUSY setting, inflation can only occur if SUSY is 
broken somehow. The purpose of the present paper is to investigate how in 
the simple situations of 5D ${\cal N}=1$ SUSY $U(1)$ gauge dynamics 
SUSY breaking translate into a 
potential for $\Theta $. We assume the extra dimensional coordinate 
$y$ to describe a circle. 
Our results can be used as 
guidelines for the more realistic cases of non Abelian gauge symmetry 
and/or a larger number of (orbifolded) extra dimensions. In the next section 
we introduce our model. We use the formulation of \cite{MartiPomarol} with a 
chiral radion field ${\cal T}$ and consider spontaneous SUSY breaking by a no-scale 
sector such that $F_T\not=0$. We also consider $D$-term SUSY 
breaking by assuming that the neutral scalar component of the chiral 
superfield $\Phi $ 
in the 5D gauge supermultiplet obtains a profile along the extra 
dimension. The effect on the spectrum of the KK modes 
is investigated for both cases. In Sections 3 and 4 we compute the 
effective one-loop potentials for $\Theta $. 
In Sec.\,5 we present cosmological applications of the models. 
Sec.\,6 has a discussion and an outlook. Appendix A is devoted to 
an analysis of the KK spectra. It also contains technical 
details concerning the computation of the effective potential. In 
Appendix B a calculation of the effective potential is presented which uses the 
world-line method.

\section{The Model}

\subsection{Set-up of 5D dynamics}

Consider 5D ${\cal N}=1$ SUSY $U(1)$ gauge theory formulation \cite{Wacker} 
with the fifth spatial and flat dimension $y$ compactified on 
a circle, $0\le y \le 2\pi R$. We stick to the usual convention that 4D 
coordinates are labeled by Greek indices 
($\mu=0,\cdots,3$). 5D ${\cal N}=1$ supersymmetry   
is equivalent to 4D ${\cal N}=2$ SUSY. In terms of 
${\cal N}=1$ supermultiplets, the 5D gauge supermultiplet is 
${\bf V}_{{\cal N}=2}=({\bf V}, \Si )$, where
\eqb
{\bf V}=-\te \si^\mu\bar{\te }A_\mu+{\rm i}\bar{\te }^2\te \lam_1-
{\rm i}\te^2\bar{\te }\bar{\lam }_1+\fr{1}{2}\bar{\te }^2\te^2D~
\label{SiV}
\eqe
is a 4D vector superfield, while 
$\Si $ is a chiral superfield
\beq
\Si =\fr{1}{\sq{2}}(\Phi +{\rm i}A_5)+\sq{2}\te \lam_2+\te^2 F_{\Si }~.
\label{defSi}
\eeq 
The 5D Lagrangian ${\cal L}_{\bf V}$ of the pure gauge sector reads 
\cite{MartiPomarol}, \cite{Wacker}
\beq
{\cal L}_{\bf V}=\fr{1}{g^2}\int d^4\te 
\left [ (\sqrt{2}\pl_5{\bf V}-\Si^+)
(\sqrt{2}\pl_5{\bf V}-\Si)-(\pl_5{\bf V})^2\right ]
+\fr{1}{4g^2}\l \int d^2\te W^{\al }W_{\al}+{\rm h.c.}\r\,,
\label{LV}
\eeq
where
$$
W_{\al }=-{\rm i}\lam_{1 \al }+
[\de_{\al }^{\bt }D-
{\rm i}(\si^{\mu \nu})_{\al }^{\bt }A_{\mu \nu}]\te_{\bt }+ 
\te^2(\si^{\mu }\pl_{\mu }\ov{\lam })_{\al }~,
$$
\beq
\si^{\mu \nu }=\fr{1}{4}(\si^{\mu}\bar \si^{\nu}-\si^{\nu}\bar \si^{\mu})~,~~~
A_{\mu \nu}=\pl_{\mu }A_{\nu }-\pl_{\nu }A_{\mu}~,~~~ 
W^{\al }W_{\al}|_{\te \te }=D^2+\cdots ~.
\label{W}
\eeq
In addition, there are two chiral superfields 
\beq
\phi =\vp  +\sq{2}\te \lam_{\phi }+\te^2F_{\phi }~,~~~
\bar{\phi }=\bar{\varphi } +\sq{2}\te \lam_{\bar{\phi }}+
\te^2F_{\bar{\phi }}
\label{phi}
\eeq
with respective charges $q$ and $-q$. These superfields constitute 
the 5D hypermultiplet 
${\bf \phi }_{{\cal N}=2}=(\phi, \bar{\phi })$. 
The 5D Lagrangian ${\cal L}_\phi$ for the matter 
reads
\beq
{\cal L}_{\phi }=\int d^4\te \l \phi^+e^{-2q{\bf V}}\phi +
\bar{\phi}e^{2q{\bf V}}\bar{\phi}^+\r
+\l \int d^2\te \bar{\phi } (M+\pl_5-\sqrt{2}q\Si )\phi 
+{\rm h.c.}\r\, 
\label{Lphi}
\eeq
where $M$ denotes a real SUSY bulk mass. 
Each of the Lagrangians (\ref{LV}), (\ref{Lphi}) is invariant 
under 5D $U(1)$ gauge transformations
$$
{\bf V}\to {\bf V}+\fr{1}{2}(\La +\La^+)~,~~~~~~
\Si \to \Si +\fr{1}{\sq{2}}\pl_5\La ~,
$$
\beq
\phi \to e^{q\La }\phi ~,~~~~~~~\bar{\phi }\to e^{-q\La }\bar{\phi }~.
\label{5Dtrans}
\eeq
Also there is invariance under two 4D ${\cal N}=1$ SUSY transformations. These two supersymmetries are
related by an $SU(2)_R$ symmetry.
The bosonic part of ${\cal L}_{\bf V}+{\cal L}_{\phi}$, involving 4D
scalar and
auxiliary components, reads
$$
({\cal L}_{\bf V}+{\cal L}_{\phi })^B=\fr{1}{2g^2}D^2-\fr{1}{g^2}\Phi
\pl_5 D+
\fr{1}{g^2}|F_{\Si }|^2+
|F_{\phi }|^2+|F_{\bar{\phi }}|^2-
qD(|\vp |^2-|\bar{\vp }|^2)+
$$
\beq
\left [ F_{\bar{\phi
}}(M\hs{-0.1cm}+\hs{-0.1cm}\pl_5-\hs{-0.1cm}q(\Phi+iA_5))\vp +
\bar{\vp }(M\hs{-0.1cm}+\pl_5-q(\Phi+iA_5))F_{\phi
}\hs{-0.1cm}-\hs{-0.1cm}
\sq{2}q\bar{\vp }F_{\Si}\vp +{\rm h.c.}\right ].
\label{LB}
\eeq
Eliminating $F$ and $D$-terms from (\ref{LB}), the part ${\cal V}$ 
in (\ref{LB}), which does not contain 4D derivatives, reads
$$
{\cal V}=\fr{1}{2g^2}(\pl_5\Phi )^2+
\fr{g^2}{2}q^2(|\vp |^2-|\bar{\vp }|^2)^2+
2g^2q^2|\vp |^2|\bar{\vp }|^2+
$$
\beq
(M-q\Phi )^2(|\vp |^2+|\bar{\vp }|^2)+
|(\pl_5-{\rm i}qA_5)\vp |^2+
|(\pl_5+{\rm i}qA_5)\bar{\vp }|^2~.
\label{pot}
\eeq
If we assume the VEV $A_5$ to be constant in $y$, the Wilson-line 
phase $\Theta$ generated by a line integration 
along the entire extra dimension reads
\eqb
\label{Wilson}
\Te=2\pi R A_5\,.
\eqe
In order to generate a 4D effective one-loop potential for the field $\Te$ by the Hosotani mechanism 
\cite{Hosotani} SUSY must be broken in some way. We distinguish two cases 
of SUSY breaking in the following: (i) spontaneous 
breaking by the dynamics of a no-scale sector, (ii) $D$-term breaking 
by a non vanishing 
(initial) value of the VEV of $\Si $'s scalar component $\Phi $ in
(\ref{defSi}).  

We will discuss the physics being potentially responsible for situation (ii) below. As for case (i) 
some technical remarks are in order. 
A superspace formulation of the theory 
${\cal L}_{\bf V}+{\cal L}_{\phi }$ involving 
a chiral radion superfield 
\eqb
\label{radion}
{\cal T}={\cal R}+iB_5+\theta\Psi^5_R+\theta^2{\cal F_T}
\eqe
has been proposed in ref.\,\cite{MartiPomarol}. The field ${\cal T}$ 
appears in terms of a 
factor $({\cal T}^\dagger+{\cal T})/(2R)$ for the first term in (\ref{Lphi}), and in terms of a 
factor $2R({\cal T}+{\cal T}^\dagger)^{-1}$ and a factor ${\cal T}/R$ for the first and 
second term, respectively, in (\ref{LV}). Upon $\theta$ and $\bar{\theta}$
integration, the elimination of the
auxiliary fields by their equations of motion 
this formulation yields the usual component-field 5D Lagrangian ${\cal L}_V+{\cal L}_{\phi}$. 
The radion field ${\cal T}$ is of the no-scale type \cite{MartiPomarol} and has a flat potential 
in the $\la {\cal T}\ra$ direction in string-derived \cite{String} supergravity models \cite{Dudas}. 
The superfunction
$G=-3\ln\,K+\ln\,|g|^2$, $K$ denoting the K\"ahler potential and $g$ the superpotential, 
in such `no-scale' \cite{Dudas} ('non minimal' \cite{Chang} or 'minimal
$SU(n,1)$' \cite{Ellwanger}) 
models can be written in the 
form $G=-3\ln\left({\cal T}+{\cal T}^\dagger-|\phi_i|^2\right)+\ln|g_3(\phi_i,{\cal T})|^2$ where $\phi_i$
 denote chiral fields, $g_3$ is a superpotential, 
 and all fields are in Planck units. $G$ is invariant under particular K\"ahler transformations being 
related to dilatations. On tree-level no-scale models have $\la {\cal R}\ra=R$ as a sliding scale in a flat potential, 
and the scale of SUSY breaking $\la {\cal F_T}\ra\equiv F_T$ is also not fixed. 
We assume $R$ 
to be stabilized by some additional dynamics, see for example \cite{Tytgat}, and appeal to the radiative
stabilization of $F_T$ as in \cite{Gersdorffgrav}. 
With ref.\,\cite{MartiPomarol} 
we will discuss in the next section how $F_T\not=0$ affects the 
spectrum of the charged KK modes whose fluctuations generate a 
4D one-loop effective potential for $\Theta$\footnote{The way the SUSY 
breaking $F_T$ is derived in \cite{MartiPomarol} using a `compensator field', which drops out of $G$, 
should be compared in detail with the methods in \cite{Ellwanger}. 
A radiatively stabilized VEV $F_T\not=0$ as it arises in gauged SUGRA 
\cite{Gersdorffgrav} was mimicked by a flat-space no-scale sector 
in \cite{MartiPomarol}. In this model a scale $W$ enters the superpotential for a conformal
compensator chiral superfield and determines the VEV $F_T$. Stabilization is therefore already
assumed in this effective description.}.

\subsection{KK decomposition and SUSY breaking}

Since we compactify on a circle the usual consideration of 
reflection parity for orbifold compactifications does not apply. 
The charged scalar fields $\phi$ and $\bar{\phi}$ have the following 
KK decompositions
\eqb
\label{KKphi}
\vp (x, y) =\fr{1}{\sq{2\pi R}}\sum_{n=-\inf }^{+\inf }
\vp^{(n)}(x)e^{{\rm i}\fr{ny}{R}}~,~~~~
\bar{\vp}(x, y) =\fr{1}{\sq{2\pi R}}\sum_{n=-\inf }^{+\inf } 
\bar{\vp}^{(n)}(x)e^{-{\rm i}\fr{ny}{R}}~,
\eqe
while the decomposition of the real field $\Phi$ is
\beq
\Phi (x, y)=\sum_{n=1}^{\inf }\Phi_{-}^{(n)}(x)\sin \fr{ny}{R}+
\sum_{n=0}^{\inf }\Phi_{+}^{(n)}(x)\cos \fr{ny}{R}~.
\label{exp}
\eeq
A similar decomposition as in (\ref{KKphi}) and (\ref{exp}) holds for the charged fermions 
$\lambda_\phi$, $\lambda_{\bar{\phi}}$ and the field $A_5$, 
respectively. In case (i) we consider spontaneous SUSY breaking. 
To assume in (\ref{exp}) 
$\Phi_{-}^{(n)}(x)\equiv 0$ and $\Phi_{+}^{(n)}(x)=\delta_{n0} f(x)$ with some slowly 
varying function $f$ would simply shift the SUSY mass $M$ by a 
finite amount. We set $\Phi_{+}^{(n)}(x)\equiv 0$ when considering case (i). 
We are interested in the generation of a 
4D effective one-loop potential for $\Te$. Since the gauginos $\lambda_{1,2}$ do not couple 
to $A_5$ they are irrelevant for the Hosotani mechanism. As in (\ref{Wilson}) we assume 
$A_5$ to be constant in $y$ -- only its 
lowest KK mode contributes. The shift of the masses $m_n^{\phi,\bar\phi}$ of the charged, bosonic 
KK modes can be calculated after a diagonalization of the ($2\times2$)-mass 
matrix for $\bar{\phi}_n$ and $\phi_n$ at the $n$th KK-level 
\cite{MartiPomarol}:
\eab
\label{KKshift}
m_n^{\vp}=m_n^{\bar{\vp}}&=&\left[\left(\frac{n}{R}\right)^2+M^2\right]^{1/2} \to 
m_n^{\vp ,\bar{\vp},\pm}=\left[\frac{1}{R^2}\l n\pm
\fr{F_T}{2}\r^2+M^2\right]^{1/2}\,,\nonumber\\ 
&&\ \ \ \ \ \ \ \ \ \ \ \ \ \ \ \ \ \ \ \ \ \ \ \ \ \ \ \
 \ \ \ \ \ \ \ \ \ \ \ \ \ \ \ \ \ \ \ \ \ (-\infty<n<\infty)\,. 
\eae
The KK masses of charged fermions,
\eqb
\label{KKferm}
m_n^{\lam}=m_n^{\bar{\lam}}=\left[\left(\frac{n}{R}\right)^2+M^2\right]^{1/2}\,,
\eqe
are not shifted. This breaking of SUSY occurs through the non zero 
$F_T$ of the radion supermultiplet (\ref{radion}).
Therefore, the SUSY transformation for the $\Psi^5_R$ state
(the fifth component of right handed gravitino) is 
$\de_{\xi }\Psi^5_R\sim \xi F_T$ ($\xi $ is a Grassmann
variable of the SUSY transformation).
Thus, $\Psi^5_R$ emerges as a goldstino,
becomes the longitudinal mode of 4D gravitino and gives mass to it, see \cite{Gersdorffgrav}.

Let us now consider case (ii). Here SUSY breaking proceeds by 
a non vanishing scalar
VEV of $\pl_5\Phi$ and therefore, the latter appears as an order parameter
for SUSY breaking. One can see from (\ref{LB}) that, 
$\lan D\ran =-\lan \pl_5 \Phi \ran$ 
(since $\lan \vp \ran =\lan \ov{\vp } \ran=0$) and $\lan
\pl_5 \Phi \ran\not=0$ indicates complete SUSY breaking through the
$D$-term. Recalling the 4D ${\cal N}=1$ SUSY transformation for the 
gaugino
\beq
\de_{\xi }\lam_1={\rm i}D\xi +\si^{\mu \nu}A_{\mu \nu }\xi ~,
\label{gauginoTrans}
\eeq
we see that in this case the $\lam_1$ gaugino appears as a goldstino.
 For simplicity we assume in (\ref{exp}) 
\eqb
\label{nondphi}
\Phi_{-}^{(n)}(x)\equiv \delta_{n1} V\,, \ \ \ \ \ \Phi_{+}^{(n)}(x)\equiv 0\,.
\eqe
This induces next and next-to-next neighbor 
interactions in the fluctuating KK tower, see Appendix A. 
Correspondingly, higher KK modes of $\Phi$ would 
cause non vanishing elements of the KK mass matrix 
further away from the diagonal. The field expectation value 
$V$ is not dynamical.\footnote{On the one hand, slow-roll for $V$ 
due to 4D tree-level dynamics ($V^2$ potential) would 
require $V\gg M_P$ which is precisely what 
we would like to avoid. On the other hand, loop effects, which 
qualitatively change the tree-level potential for $V$ such that 
$V$ is slowly rolling at $V<M_P$, are 
not under control in a perturbative treatment of the supertrace 
appearing in the computation 
of the Hosotani potential for $\Theta$.}. For the one-loop potential 
${\cal V}^{\tiny{\mbox{eff}}}(\Theta)$ to be relevant for inflation, 
we assume here that some external dynamics keeps 
$V(x)$ constant and away from zero and that the overall vacuum energy 
vanishes at tree-level. This would happen, for example, if a SUSY-breaking 
potential $P(V)$ was added to the effective 4D dynamics 
such that the sum of the tree-level potential $\sim R^{-2}V^2$ and $P(V)$ has a 
nontrivial minimum at zero energy. 
Since only very particular potential couplings are allowed by 5D SUSY, one
can induce $P(V)$ on the 4D level. This would require an orbifold
scenario with appropriate brane potential couplings. The calculation of the
effective potential for a Wilson-line phase within an orbifold 
construction is beyond the scope of this
paper \cite{ourprep}. 
    
\section{${\cal V}^{\rm eff}(\Te )$ from $F$-term SUSY breaking}

At tree-level the Wilson-line phase $\Theta$ would not have a potential in 4D due 
to 5D gauge invariance. For a compactified extra dimension 
this is no longer true if the quantum fluctuations 
of charged bulk matter are integrated out \cite{Hosotani}. We assume 
that w.r.t. 4D-$x$ $\la A_5\ra$ varies on scales 
larger than the compactification scale $R^{-1}$ such that the lowest order 
in a derivative expansion of the quantum effective 
potential in 4D is a good approximation. 

When calculating the one-loop effective potential in 4D for $\Theta$ 
the presence of the $A_5$ coupling to 
$\vp^{(n)},\ov{\vp }^{(n)}$ and 
$\lambda_{\phi }, {\lambda_{\bar{\phi }}}$ in (\ref{LB}) and the fact that
$F_T\not=0$ 
effectively modify the KK masses as follows
\eab
\label{thetKKmass}
m_n^{\vp,\bar{\vp},\pm}&=&
\left[\frac{1}{R^2}\l n+\frac{q\Theta }{2\pi }\pm
\fr{1}{2}F_T\r^2+M^2\right]^{1/2}\,,\nonumber\\ 
m_n^{\lam}=
m_n^{\bar{\lam}}&=&\left[\frac{1}{R^2}\l 
n+\frac{q\Theta}{2\pi}\r^2+M^2\right]^{1/2}\,.
\eae
After a Schwinger parametrization of tr log we have 
\eab
\label{VeffFT}
{\cal V}^{\tiny{\mbox{eff}}}(\Theta)&=&-\frac{1}{(4\pi)^2}
{\cal S}{\mbox{Tr}}\int_0^\infty \frac{dt}{t^3}\exp(-{\cal M}t)\nonumber\\ 
&=&-\frac{1}{(4\pi)^2}\sum_{n=-\infty}^{\infty}\int_0^\infty
\frac{dt}{t^3}\exp(-M^2t)\,
\left\{\exp\left[-\fr{t}{R^2}\l 
n+\fr{q\Te }{2\pi }+\fr{F_T}{2}\r^2\right ]\,+\right.\nonumber\\ 
& &\left.\exp \left[-\fr{t}{R^2}\l
n+\fr{q\Te }{2\pi }-\fr{F_T}{2}\r^2 \right]-
2\exp\left[-\fr{t}{R^2}\l n+\frac{q\Theta}{2\pi} \r^2\right]\right\}\,,
\eae
where ${\cal S}{\mbox{Tr}}$ and ${\cal M}$ denote the supertrace and the 
(diagonal) super mass matrix for the fluctuating 
bosonic and fermionic fields, respectively. As usual\footnote{See, however, 
Appendix B 
where we sketch a calculation of ${\cal V}^{\tiny{\mbox{eff}}}(\Theta)$ 
using the world-line formalism.}, we perform a 
Poisson resummation, 
\beq
\sum_{n=-\inf }^{+\inf}e^{-\al^2 (n+\si)^2}=\fr{\sq{\pi }}{\al }
\sum_{k=-\inf }^{+\inf}e^{-\fr{\pi^2k^2}{\al^2}}e^{2{\rm i}\pi k\si }~,
\label{pois}
\eeq
and afterwards carry out the $t$ 
integral in (\ref{VeffFT}). The result of the integration is 
$2\,M^5\,K_{5/2}(2\pi RM k)$ where $K_{5/2}$ denotes a modified Bessel function. 
Collecting everything, we finally obtain 
\eab
\label{Veffresult}
{\cal V}^{\tiny{\mbox{eff}}}(\Theta)&=&\frac{3}{16\pi^6}\frac{1}{R^4}
\sum_{k=1}^{\infty}\exp(-2\pi k\,RM)\,[1-\cos(\pi k\,F_T)]\times\nonumber\\ 
& &\frac{\cos(qk\Theta)}{k^5}\left[1+2\pi k\,RM+\frac{1}{3}(2\pi
k\,RM)^2\right]\, ,
\eae
in complete agreement with a world-line calculation (Appendix B).
Notice that the divergent bosonic and fermionic 
contributions at $k=0$ cancel exactly because of SUSY - the usual 
fine-tuning of the cosmological constant is not needed. 
For $F_T\not=0$ and small $\Theta$ the effective potential 
is positive, and thus it drives de Sitter expansion if $\Theta$ is rolling 
sufficiently slowly. In the limit of unbroken SUSY, 
$F_T\to 0$, the effective one-loop potential 
for $\Theta $ vanishes. 

\section{${\cal V}^{\rm eff}(\Te )$ from $D$-term SUSY breaking}

Let us now investigate case (ii), namely the effect of $D$-term SUSY
breaking 
according to (\ref{nondphi}). The one-loop effective potential in this case reads
\beq
{\cal V}^{\rm eff}=-\fr{1}{16\pi^2 }{\cal S}{\rm Tr}
\int_0^{\inf }\fr{dt}{t^3}\left[e^{-(D+B)t}\right]~,
\label{pot1}
\eeq
where ${\cal S}{\rm Tr}$ also indicates the trace in KK space. 
The matrices $D$ and $B$ denote the diagonal and
the off-diagonal parts of the KK mass-squared matrices. 
We will compute ${\cal V}^{\rm eff}$ in quadratic order in $B$. 
The treatment of the term 
$\fr{(-1)^l}{l!}t^l{\cal S}{\rm Tr}(D+B)^l$ [expansion of the exponential in (\ref{pot1})] 
is presented in Appendix A. Under consideration of 
(\ref{partsum}), the quadratic order in $B$ reads
$$
{\cal S}{\rm Tr}\,(e^{-(D+B)t})=-2\l \fr{qV}{2R}\r^2
\sum_{n=-\inf}^{+\inf} \sum_{l=2}^{+\inf}\fr{(-1)^l}{(l-1)!}t^l
\fr{D_n^{l-1}-D_{n-1}^{l-1}} {D_n-D_{n-1}}=
$$
\beq
2\l \fr{qV}{2R}\r^2 t
\sum_{n=-\inf}^{+\inf} \fr{e^{-D_nt}-e^{-D_{n-1}t}} {D_n-D_{n-1}}=
-2\l \fr{qV}{2R}\r^2 t^2\sum_{n=-\inf}^{+\inf}
\int_0^1dy\,e^{-D_n(1-y)t-D_{n-1}yt}~.
\label{StrTr}
\eeq
We now perform a Poisson resummation (\ref{pois}) 
in (\ref{StrTr}). Recalling that
\beq
D_n=\fr{1}{R^2}\l n-\fr{q\Te }{2\pi }\r^2+M^2+\fr{q^2V^2}{2}~,
\label{defDn}
\eeq
we obtain
\beq
{\cal V}^{\rm eff}=\fr{1}{8\pi^{5/2}}\l \fr{qV}{2R}\r^2R\int_o^1dy
\sum_{k=-\inf }^{+\inf }e^{-{\rm i}kq\Te -2{\rm i}\pi ky}I_k~
\label{efpot1}
\eeq
with
\beq
I_k=\int_{0}^{\inf }\fr{dt}{t^{3/2}}{\rm exp}
\lq -\l M^2+\fr{q^2V^2}{2}+R^{-2}y(1-y)\r t-\fr{\pi^2R^2}{t}k^2\rq\,.
\label{Ik}
\eeq
Performing the $t$-integration for $|k|\ge 1$, we get
\beq
I_k=\fr{1}{\sq{\pi }R|k|}{\rm exp}
\lq -2\pi R|k|\l M^2+q^2V^2/2+R^{-2}y(1-y)\r^{1/2}\rq ~.
\label{Ik1}
\eeq
The case $k=0$ leads to a $\Te $-independent, linear UV divergence 
which renormalizes the 4D gauge coupling (a discussion of this is presented 
at the end of the paper). We omit this $\Theta$ independent part 
when studying inflation. Thus the final expression for 
${\cal V}^{\rm eff}$ reads
\beq
{\cal V}^{\rm eff}=\fr{1}{4\pi^2}\l \fr{qV}{2R}\r^2
\sum_{k=1}^{+\inf }\fr{\cos (qk\Te )}{k}
~{\cal C}_k~,
\label{finefpot}
\eeq
where
\beq
{\cal C}_k=
\int_0^1dy \cos (2\pi ky)\,
{\rm exp} 
\lq -2\pi Rk\l M^2+q^2V^2/2+R^{-2}y(1-y)\r^{1/2}\rq 
~.
\label{Ck}
\eeq
The same result is obtained by a (very fast!) world-line method calculation
in Appendix B.
Notice the exponential suppression of higher winding modes in the 
coefficient ${\cal C}_k$. 
If there is a hierarchy $MR\gg 1$ the effective 
potential is strongly suppressed.

The effective potentials in (\ref{Veffresult}), (\ref{finefpot}) both are
invariant under
transformations $\Te \to \Te+\fr{2\pi }{q}m$ ($m$ integer). This is
an expected result. Due to the form of KK mass spectrum in 
(\ref{thetKKmass}), (\ref{scMatrel}), (\ref{ferMatrel}), 
the Lagrangian is not changing under this shift
because there is summation over an infinite number of KK states. This
manifests the 5D gauge invariance. 

\section{Cosmological applications}

\subsection{Slow roll and ${\rm e}$-foldings during inflation}

To decide under what conditions 
$\Theta$ can drive inflation we 
have to look at its 4D kinetic term as it follows from 
the 5D gauge curvature by integrating over $y$ \cite{AHRandall}. 
This may seem strange at first sight since we used a derivative 
expansion in zeroth-order approximation when calculating 
${\cal V}^{\tiny{\mbox{eff}}}(\Theta)$. 
However, in the slow-roll regime derivative terms arising from the expansion of 
${\cal V}^{\tiny{\mbox{eff}}}(\Theta,\pd_\mu\Theta)$ 
are strongly suppressed as compared to the ones coming from 
the 5D gauge curvature, see below. We have
\eqb
\label{4Dkan}
{\cal L}^{4D}_\Theta=\frac{1}{2}
\frac{\pd^\mu\Theta\pd_\mu\Theta}{(2\pi R)^2g_4^2}-
{\cal V}^{\rm eff}(\Theta)\,.
\eqe
So the canonically normalized field $\phi_\Theta$ is defined as
\eqb
\label{cannorm}
\phi_\Theta\equiv f_\Theta\Theta\equiv \frac{\Theta}{2\pi R g_4}\,,
\eqe
where $(g^{(4)})^2\equiv \frac{(g^{(5)})^2}{2\pi R}$ and $M_P$ is the 4D reduced Planck mass. 
The slow-roll conditions 
for the field $\phi_\Theta$ read
\eqb
\label{sr}
\ep=\frac{(M_P)^2}{2}\left(
\frac{({\cal V}^{\rm eff})'}{{\cal V}^{\rm eff}}\right)^2\ll 1\,,\ \ \ 
|\eta|=(M_P)^2\left|
\frac{({\cal V}^{\rm eff})''}{{\cal V}^{\rm eff}}\right|\ll 1\, ,
\eqe
where $({\cal V^{\rm eff}})'$, $({\cal V^{\rm eff}})''$
denote the potential's derivatives in respect
of $\phi_\Theta $.
Let us first discuss case (i). We only consider the contribution 
at $k=1$ in (\ref{Veffresult}) since terms with $k>1$ are 
exponentially {\sl and} strongly power suppressed. Substituting the $k=1$ part 
of (\ref{Veffresult}) or (\ref{finefpot}) into (\ref{sr}), we obtain
\eab
\label{sr(i)}
\ep&=&\frac{1}{2}\,(2\pi q g_4)^2\,(M_P R)^2\,\tan^2[2\pi q g_4 R\phi_\Theta]\nonumber\\ 
|\eta|&=&(2\pi q g_4)^2\,(M_P R)^2\,. 
\eae
According to (\ref{sr(i)}) a hierarchy between $R^{-1}$ and $M_P$ can be compensated 
by a small 4D gauge coupling $g_4\ll 1$. Moreover, a small $g_4$ keeps the argument 
of the tangent small in $\ep$ and thus yields an additional suppression. 
Let us now look at the amount of ${\rm e}$-foldings we may expect to be
produced 
by the slowly rolling field $\phi_\Theta$.  

The number of ${\rm e}$-foldings for canonically normalized field
$\phi_\Theta $
(of $\Te $) is given by
the following formula
\beq
N_{\Te }=\int_{t^i}^{t^f}Hdt\simeq \fr{8\pi }{M_P^2}
\left |\int_{\phi^i_\Te }^{\phi^f_\Te } 
\fr{{\cal V^{\rm eff}}}{({\cal V^{\rm eff}})'}d \phi_\Theta \right |~.
\label{efoldings}
\eeq
Using (\ref{Veffresult}),  
(\ref{finefpot}),  (\ref{cannorm}), for both
scenarios considered above, we obtain 
\beq
N_{\Te }=\fr{\rho }{(M_PR)^2\pi g_4^2}~,~~~~~~
\rho =\frac{2}{q^2}\ln \left |\sin (2\pi q g_4 R\phi_\Theta ) 
\right |_{\phi^i_\Te }^{\phi^f_\Te } \sim {\cal O}(1)~.
\label{efoldresult}
\eeq
Taking in (\ref{efoldresult}) $M_PR=10-100$~, 
$g_4\stackrel{<}{_\sim } 10^{-2}-10^{-3}$,
we easily get $N_{\Te }\stackrel{>}{_\sim }55$, which evades the horizon
problem and guarantee flatness of the Universe up to the needed accuracy.

\subsection{Application for quintessence}

As was pointed out in \cite{Riotto},
for a hierarchy $1/R\ll M$
(recall that $M$ is a SUSY bulk mass for matter), the generated effective
potentials are getting strongly suppressed. From 
(\ref{Veffresult}), (\ref{finefpot}) 
one sees that the potentials in both scenarios (i) and (ii) aquire a factor
$\sim \exp\l -2\pi MR\r $, which for $MR=30-50$ provides suppression 
such that 
$\lan {\cal V}^{\rm eff}\ran \sim (3\cdot 10^{-3}~{\rm eV})^4$.
This amount can be associated with dark energy, making the model a candidate
for quintessence \cite{quint}.

The presented effective potentials may be relevant for both 
inflation and quint\-essence. Infla\-tion would start at $RM\sim 1$, 
and at the end of inflation or after reheating the radion field would relax to the minimum of a stabilizing potential 
correspon\-ding to $MR\gg 1$. This, of course, requires a dynamical model 
for the radion field, for its foundation within 5D supergravity see \cite{Gersdorffgrav,Kugo}.


       
\section{Discussion and outlook}

In this paper we have realized the idea of gauge inflation within a 5D
supersymmetric setting exploiting $U(1)$ gauge symmetry. For this, two
possibilities of SUSY breaking were considered. The 'no scale'
one-loop potential has the advantage that the tree-level contribution vanishes. 
In both scenarios, the obtained effective
potentials make the 'Wilson-line
phase $\Theta $' a candidate for the inflaton. 
Namely, for a given hierarchy between $M_P$ and the compactification scale
$1/R$ a proper choice of the 4D gauge coupling $g_4$ guarantees slow 
roll conditions and the required number of ${\rm e}$-foldings. Due to SUSY the fine-tuning of the 4D
cosmological constant to zero after inflation is with respect to $\sim R^{-4}$ as compared 
to the the non-SUSY value of $\sim R(M_P^{(5)})^5$ where $M_P^{(5)}\sim M_P$ 
denotes the 5D Planck scale.
Also, the specific forms of the effective potentials open up the possibility to describe 
quintessence. 

As a byproduct we have demonstrated that the winding mode representation
can be written down directly using the world-line method (Appendix B).

For quantitative estimates [in
eqs. (\ref{sr})-(\ref{efoldresult})], one should use a perturbatively small, 
renormalized value of $g_4$. As it is known the KK excitations above $1/R$ induce a power law 
running of $g_4$ (linear in case of 5D). This effect was observed 
in our calculation of ${\cal V}^{\rm eff}(\Te )$ in the case of
$D$-term SUSY breaking. The emerging linear UV divergence, coming from the zero
winding mode, should be absorbed into the bare $1/g_4^2$ [the corresponding
operator is $\fr{1}{g_4^2}(\pl_5 \Phi )^2$]. This is nothing but a 
renormalization of the gauge coupling. Another question, which should be
understood (for both scenarios), is the origin of the small value
for $g_4$ and the mechanism which guarantees its smallness. 
For example, an embedding of the $U(1)$ in a non Abelian gauge symmetry, which is unbroken 
at high energies, prevents $g_4$ to reach the Landau pole. For such a scenario, one could 
have in mind some specific GUT, which naturally unifies $U(1)$ with the SM
interactions. Another way for an ultraviolet completion is the
possibility of embedding the 5D SUSY $U(1)$ gauge theory in a superstring 
model.

\vspace{0.3cm}
\hs{-0.6cm}{\bf Acknowledgements}

\hs{-0.6cm}Research of F.P.C. is supported by 
Funda\c c\~ ao para a Ci\^ encia e a 
Tecnologia 
(grant SFRH/BD/4973/2001). 

\vspace{0.2cm}


\section*{Appendix A: KK spectrum and ${\cal S}\mbox{Tr}\left[\e^{-(D+B)t}\right]$
} 
\setcounter{equation}{0} 
\renewcommand{\theequation}{A.\arabic{equation}} 

Here we quote the mass matrices for the fluctuating KK modes as they arise 
from the $D$-term SUSY breaking according to (\ref{nondphi}). 
After KK decomposition of the fields appearing in 
(\ref{pot}) and subsequent integration over $y$ 
the bilinear terms in the effective 4D bosonic KK modes are
\eqb
\label{massspecKK}
{\cal V}^{(2)}=\sum_{n,m=-\inf }^{+\inf }\left\{\vp^{(n)}M^2_{m,n}\vp^{(m)*}+
\bar{\vp }^{(n)}\bar{M}^2_{m,n}\bar{\vp }^{(m)*}\right\}\,.
\eqe
The bilinear expressions for the superpartners have the form
\beq
-{\cal L}^{(2)}_{\lam }=\sum_{n,m=-\inf }^{+\inf }\left\{ 
\lam_{\bar{\phi }}^{(n)}M^{\lam }_{n,m}\lam_{\phi }^{(m)}+
\ov{\lam }_{\bar{\phi }}^{(n)}
M^{\lam *}_{n,m}
\ov{\lam }_{\phi }^{(m)}\right\}~.
\label{fer2}
\eeq
The matrix elements appearing in (\ref{massspecKK}) and (\ref{fer2}) read 
as follows
\eab
\label{scMatrel}
M^2_{n,n}&=&\bar{M}^2_{n,n}=\l \fr{n}{R}-\fr{q\Te }{2\pi R}\r ^2
+M^2+\fr{q^2}{2}V^2~,~~~\nonumber\\ 
M^2_{n,n+1}&=&-M^2_{n,n-1}=
-\bar{M}^2_{n,n+1}=\bar{M}^2_{n,n-1}={\rm i}qMV~,\nonumber\\ 
M^2_{n,n+2}&=&M^2_{n,n-2}=\bar{M}^2_{n,n+2}=\bar{M}^2_{n,n-2}=-\fr{q^2}{4}V^2~,
\eae
all other elements are zero. The matrix elements of fermionic states are
\beq
M^{\lam }_{n, n}=
{\rm i}\l \fr{n}{R}-\fr{q\Te }{2\pi R}\r +M~,~~~
M^{\lam }_{n, n-1}=-M^{\lam }_{n, n+1}=
\fr{{\rm i}q}{2}V~,
\label{ferMatrel}
\eeq
while all other elements are zero. For the calculation of the effective potential the matrix 
$M^{\lam }M^{\lam +}\equiv (M^2)^{\lam }$ is relevant where $M^{\lam +}$ denotes 
complex conjugation of $M^{\lam }$ and transposition in the
KK indices. The elements of $(M^2)^{\lam }$ are
\eab
(M^2)^{\lam }_{n, n}&=&\l \fr{n}{R}-\fr{q\Te }{2\pi R}\r^2+M^2+
\fr{q^2}{2}V^2~,~~
(M^2)^{\lam }_{n, n+2}=(M^2)^{\lam }_{n+2, n}=-\fr{q^2}{4}V^2~,\nonumber\\ 
(M^2)^{\lam }_{n, n+1}&=&-{\rm i}qMV-\fr{qV}{2R}~,~~~
(M^2)^{\lam }_{n+1, n}={\rm i}qMV-\fr{qV}{2R}\,.
\label{ferMatr2el}
\eae
All other elements are zero. 

Let us now evaluate the expression 
\beq
{\cal S}{\rm Tr}e^{-(D+B)t}={\cal S}{\rm Tr}\sum_{l=0}^{+\inf}\fr{(-1)^l}{l!}t^l(D+B)^l~.
\label{expexpon}
\eeq
We have 
\beq
(D+B)^l=D^l+\sum_{p=0}^{l-1}D^{l-1-p}BD^p+
\sum_{p=0}^{l\hs{-0.1mm}-2}
\sum_{q=0}^{l\hs{-0.1mm}-\hs{-0.1mm}2\hs{-0.1mm}-\hs{-0.1mm}p}
D^{l-2-p-q}BD^pBD^q+\cdots\,. 
\label{power}
\eeq
The supertrace of the first term in (\ref{power}) vanishes because the elements of $D$ coincide for
scalars and fermions, see (\ref{scMatrel}) and (\ref{ferMatr2el})]. 
The trace of the second term in (\ref{power}) is zero because $B$ is
an off-diagonal matrix. Therefore the leading contributions are due to the third 
term in (\ref{power}). We have
$$
{\rm Tr}\l D^{l-2-p-q}BD^pBD^q\r ={\rm Tr}\l D^{l-2-p}BD^pB\r =
$$
\beq
\sum_{n=-\inf}^{+\inf}\l D^{l-2-p}_nD^p_{n\pm 1}|B_{n, n\pm 1}|^2+
D^{l-2-p}_nD^p_{n\pm 2}|B_{n, n\pm 2}|^2\r
\label{Tr}
\eeq
\beq
{\cal S}{\rm Tr}\l D^{l-2-p-q}BD^pBD^q\r =
-2\l \fr{qV}{2R}\r^2
\sum_{n=-\inf}^{+\inf}D^{l-2-p}_n\l D^p_{n-1}+D^p_{n+1}\r~.
\label{STr}
\eeq
According to (\ref{power}), (\ref{STr}) we evaluate the following sum
$$
\sum_{n=-\inf}^{+\inf}\sum_{p=0}^{l\hs{-0.1mm}-2} 
\sum_{q=0}^{l\hs{-0.1mm}-\hs{-0.1mm}2\hs{-0.1mm}-\hs{-0.1mm}p} 
D^{l-2-p}_n\l D^p_{n-1}+D^p_{n+1}\r=
\sum_{n=-\inf}^{+\inf}\sum_{p=0}^{l\hs{-0.1mm}-2} 
(l-1-p)D^{l-2-p}_n\l D^p_{n-1}+D^p_{n+1}\r =
$$
$$
\sum_{n=-\inf}^{+\inf}\sum_{p=0}^{l\hs{-0.1mm}-2} 
(l-1-p)D^{l-2-p}_n D^p_{n-1}+
\sum_{n=-\inf}^{+\inf}\sum_{p=0}^{l\hs{-0.1mm}-2} 
(1+p)D^{l-2-p}_n D^p_{n-1}=
$$
\beq
\sum_{n=-\inf}^{+\inf}
\sum_{p=0}^{l\hs{-0.1mm}-2} lD^{l-2-p}_n D^p_{n-1}=
\sum_{n=-\inf}^{+\inf} l\fr{D_n^{l-1}-D_{n-1}^{l-1}}
{D_n-D_{n-1}}~.
\label{sum}
\eeq
We have obtained the second term in the second line of (\ref{sum}) by 
substituting $p\to l-2-p$ and shifting $n\to n-1$. Finally, we have
\eqb
\label{partsum}
{\cal S}{\rm Tr}(D+B)^l=
-2\l \fr{qV}{2R}\r^2\sum_{n=-\infty}^{+\infty}
\sum_{p=0}^{l\hs{-0.1mm}-2} lD^{l-2-p}_n D^p_{n-1}=
-2\l \fr{qV}{2R}\r^2\sum_{n=-\infty}^{+\infty} l\fr{D_n^{l-1}-D_{n-1}^{l-1}}
{D_n-D_{n-1}}~.
\eqe

\section*{Appendix B: Calculation of 
${\cal V}^{\tiny{\mbox{eff}}}(\Te)$ in world-line formalism}
\setcounter{equation}{0} 
\renewcommand{\theequation}{B.\arabic{equation}} 

Here we briefly sketch how an effective one-loop potential for 
$\Theta$ arises in the world-line formalism. First, we consider bosonic fluctuations. 
Like in thermal physics \cite{therm} the effective 
one-loop potential in the world-line formalism is expressed in terms of an 
integration over the proper length of all possible 
closed-path trajectories. We have
\eab
\label{WLTH}
\int d^4x\,{\cal V}^{\tiny{\mbox{eff}}}(\Theta)&=&-\int d^5x\int_0^\infty
\frac{dT}{T}\sum_{k=-\infty}^{\infty}
\int {\cal D}^5y\nonumber\\ 
& &\exp\left[-\int_0^T d\tau \left(\frac{(\dot{x}_M)^2}{4}+iq\dot{x}_5
A_5+M^2\right)\right]\,,
\eae
where
\eqb
\label{defx5}
x_M(\tau)=2\pi R\,k\frac{\tau}{T}\delta_{M5}+y_M(\tau)+x_M\,,\ \ \ \int_0^Td\tau\,y_M(\tau)=0\,.
\eqe
In (\ref{defx5}) a constant part $x_5$ and a non constant, periodic part $y(\tau)$ have 
been separated from the topological part of the trajectory. 
Performing the integration over all trajectories and using 
$\Theta=2\pi RA_5$, we arrive at

$$
\int
d^4x\,{\cal V}^{\tiny{\mbox{eff}}}(\Theta)\hs{-0.1cm}=\hs{-0.1cm}-\int
\hs{-0.1cm}d^5x\hs{-0.1cm}\sum_{k=-\infty}^{\infty}\hs{-0.2cm}\exp[ikq\Theta]
\int_0^\infty \hs{-0.1cm}\frac{dT}{T}\frac{1}{(4\pi T)^{5/2}}
\exp\left[-\frac{(\pi\,Rk)^2}{T}-M^2T\right]
$$
$$ 
\hs{-0.3cm}=\hs{-0.1cm}-\int
d^4x\,\frac{3}{16\pi^6}\frac{1}{R^4}\sum_{k=1}^{\infty}\exp(-2\pi nRM)\frac{\cos(kq\Theta)}{k^5}
 \left[1+2\pi k\,RM+\frac{1}{3}(2\pi k\,RM)^2\right]
$$
\beq
-\int d^5x\int_{1/\La^2}^{\infty }\fr{dT}{T}\fr{1}{(4\pi T)^{5/2}}
\exp\l -M^2T\r \,.
\label{WLTHinb}
\eeq 
The last term in (\ref{WLTHinb}) is a contribution from $k=0$ and is
quintically UV divergent. 
It expresses the usual cosmological-constant problem in 5D. Whether a 
cutoff must be introduced or not in the KK tower needs not be 
addressed in the world-line formulation. 
In particular in the non-SUSY case we do, however, need a 5D cutoff 
$\La^{-2}\le T$ for the zero-winding contribution. 

In case (i) (spontaneous, no-scale SUSY breaking) the coupling to the gauge field of 
the two complex, fluctuating scalars and the two Dirac fermions 
is in the world-line formalism described by

\begin{equation}
\begin{array}{cc}
 & {\begin{array}{cc}
\hspace{-7mm}  & 
  
\end{array}}\\ \vspace{2mm}
\begin{array}{c}
 \\  \\
 \end{array}\!\!\!\!\!&{\rm i}\dot{x}_5{\left(\begin{array}{ccc}
qA_5 &\, \frac{{\rm i}}{2R}F_T
\\  
-\frac{{\rm i}}{2R}F_T  &\,qA_5 
\end{array}\right) }~,~~~~~~{\rm i}q\dot{x}_5A_5~,
\end{array}  \!\!  ~~~~~
\label{sccouplg}
\eeq
%
respectively. Using these interactions to calculate 
the effective potential in analogy to (\ref{WLTH}) and taking a trace, 
we again arrive at expression (\ref{Veffresult}) which was 
obtained from a Poisson resummation of the KK spectrum. 
 
In case (ii) ($D$-term SUSY breaking by $\Phi=V\sin\frac{y}{R}$) the
world-line treatment 
is more involved since now the `outer' field $\Phi$ 
is not constant \cite{outWL}. Let us briefly sketch how things 
work here. For charged scalars we only have to shift $M\to M-\frac{q}{2}\Phi$ 
in (\ref{WLTH}). For Dirac fermions interacting with $\Phi$ the path 
integral in (\ref{WLTH}) is enlarged by a world-line Grassmann integration 
\eqb
\label{WLGR}
\int{\cal D}{{\cal \psi}_6}\,\exp(-\int_0^T\,d\tau\left(\frac{1}{2}\psi^M\dot{\psi}_M+
\frac{1}{2}\psi_6\dot{\psi}_6-
2i\psi_6\psi_5\pd_5\Phi\right)\,.
\eqe
The additional component $\psi_6$ is needed if the coupling of the spinning particle 
to the scalar background $\Phi$ is of the Yukawa-type in 
field theory \cite{outWL}. Since the part in (\ref{WLTH}), which 
depends on $x_5$ is universal 
for fermionic and bosonic fluctuations we only need to consider the term 
$qi\psi_6\psi_5\pd_5\Phi$ in (\ref{WLGR}). This term breaks the supersymmetric cancellation. 
We write $\Phi=V/(2i)(\exp[ix_5(\tau)/R]-\exp[-ix_5(\tau)/R])$, substitute 
this into (\ref{WLGR}) 
and expand the exponential in (\ref{WLGR}) up to second order in $V$. 
Performing the Grassmann-integration and 
exploiting translational invariance on the circle, the part of (\ref{WLGR}) due to 
$qi\psi_6\psi_5\pd_5\Phi$ turns into
\eqb
\label{underint}
-2T\frac{V^2}{R^2}\int_0^T d\tilde{\tau}\, G_F^2(\tilde{\tau})
\exp(-G_B(\tilde{\tau})/R^2)\cos(2\pi k\frac{\tilde{\tau}}{T})\,,
\eqe
where $G_F=\mbox{sign}\tilde{\tau}$ and $G_B=\tilde{\tau}(1-\tilde{\tau})/T$ 
denote the fermionic and bosonic world-line Green's functions, respectively.
Substituting $\tilde{\tau}=yT$ in (\ref{underint}), performing the 
$T$- and the $x^5$-integrations, and neglecting the $\Phi $ dependent 
contributions from the bosonic sector and the linearly UV divergent $k=0$ 
contribution, we arrive at the following effective potential
\eqb
\label{finwl}
\frac{(qV)^2}{16\pi^2R^2}\sum_{k=1}^\infty 
\frac{\cos(kq\Theta)}{k}\int_0^1 dy\,\cos(2\pi ky)\,\exp(-2\pi k(M^2R^2+y(1-y))^{1/2})\,.
\eqe
This is the formula one would obtain from (\ref{finefpot}) by isolating
the 
quadratic order in $V$.

\bibliographystyle{prsty}

\begin{thebibliography}{10}

\bibitem{Guth}
A.H. Guth, Phys. Rev.{\bf D}{\bf 23}, 347 (1981);\\ 
A.D. Linde, Phys. Lett.{\bf B} {\bf 108}, 389 (1982).  

\bibitem{copel}
E. Copeland et al., Phys. Rev. D 49, 6410 (1994);\\
G. Dvali, Q. Shafi, R. Schaefer, Phys. Rev. Lett. 73 , 1886 (1994). 

\bibitem{LittleHiggs}
N. Arkani-Hamed, A. G. Cohen, and H. Georgi, Phys. Rev. Lett. {\bf 86}, 4757 (2001) [hep-th/0104005];\\  
C. T. Hill, S. Pokorski, J. Wang, Phys. Rev. {\bf D64}, 105005 (2001) [hep-th/0104035];\\   
N. Arkani-Hamed, A.G. Cohen, E. Katz, A.E. Nelson, T. Gregoire, and J.G. Wacker,
JHEP {\bf 0208}, 021 (2002);\\ 
I. Low, W. Skiba, and D. Smith, Phys. Rev. {\bf D66}, 072001 (2002) 
[hep-ph/0207243]; D. E. Kaplan and M. Schmaltz, hep-ph/0302049.  

\bibitem{curvaton}
S. Mollerach, Phys. Rev. {\bf D42}, 313 (1990);\\ 
D.H. Lyth and D. Wands, Phys. Lett. {\bf B524}, 5 (2002) [hep-ph/0110002];\\ 
T. Moroi and T. Takahashi, Phys. Lett. {\bf B522}, 215 (2001) [hep-ph/0110096];\\ 
K. Dimopoulos, D.H. Lyth, A. Notari, and A. Riotto, hep-ph/0304050.     

\bibitem{Freese}
K. Freese, J.A. Friemann, and A.V. Olinto, Phys. Rev. Lett. {\bf 65}, 3233 (1990);\\ 
F.C. Adams, J.R. Bond, K. Freese, J.A. Friemann, and A.V. Olinto, 
Phys. Rev.{\bf D}{\bf 47}, 426 (1993) [hep-ph/9207245].

\bibitem{AHRandall}
N. Arkani-Hamed, H.-C. Cheng, P. Creminelli, and L. Randall, hep-th/0302034, hep-th/0301218;\\   
D. Kaplan and N. Weiner, hep-ph/0302014.

\bibitem{Zhang}
B. Feng, M. Li, R.-J. Zhang, and X. Zhang, astro-ph/0302479.

\bibitem{WMAP}
C.L. Bennett et al., astro-ph/0302207.

\bibitem{Hosotani}
Y. Hosotani, Phys. Lett.{\bf B} {\bf 126}, 309 (1983); hep-ph/0303066.

\bibitem{Riotto}
L. Pilo, D.A.J. Rayner, and A. Riotto, hep-ph/0302087. 

\bibitem{quint}
C. Wetterich, Nucl. Phys. {\bf B302}, 668 (1988);\\
B. Ratra, P.J. Peebles, Phys. Rev. {\bf D37}, 3406 (1988);\\
R.R. Caldwell, R. Dave, P.J. Steinhardt, Phys. Rev. Lett. {\bf 80}, 1582 (1998).


\bibitem{Barb}
I. Antoniadis, Phys. Lett. {\bf B246}, 377 (1990);\\ 
R. Barbieri, L.J. Hall, Y. Nomura, Phys. Rev. {\bf D63}, 105007 (2001) [hep-ph/0011311];\\ 
N. Arkani-Hamed, L.J. Hall, Y. Nomura, D.R. Smith, and N. Weiner, Nucl. Phys. {\bf B605}, 81 (2001) 
[hep-ph/0102090];\\ 
 I. Antoniadis, K. Benakli, and  M. Quiros, New J. Phys. {\bf 3}, 20.1-20.24 (2001) [hep-th/0108005];\\ 
G. von Gersdorff, N. Irges, and M. Quiros, Nucl. Phys. {\bf B635}, 127 (2002) [hep-th/0204223], hep-th/0206029. 

\bibitem{Nilles}
D.M. Ghilencea and H.-P. Nilles, Phys. Lett. {\bf B507}, 327 (2001)  [hep-ph/0103151];\\ 
D.M. Ghilencea, H.-P. Nilles, and S. Stieberger, New J.Phys.{\bf 4}, 15 (2002) [hep-th/0108183];\\ 
D.M. Ghilencea, S. Groot Nibbelink, and  H.-P. Nilles, Nucl. Phys. {\bf B619}, 385 (2001) [hep-th/0108184].   

\bibitem{WLF}
M.J. Strassler, Nucl. Phys. {\bf B385}, 145 (1992);\\ 
M.G. Schmidt and C. Schubert, Phys. Lett. {\bf B318}, 438 (1997); 
{\sl ibid} {\bf B331}, 69 (1994); Phys. Rev. {\bf D53}, 2150 (1996);\\ 
M. Reuter, M.G. Schmidt, and C. Schubert, Ann. Phys. {\bf 259}, 313 (1997). 

\bibitem{MartiPomarol}
D. Marti and A. Pomarol, Phys. Rev.{\bf D64}, 105025 (2001) [hep-th/0106256].

\bibitem{Wacker}
N. Arkani-Hamed, T. Gregoire, and J. Wacker, 
JHEP {\bf 0203}, 055 (2002) [hep-th/0101233];\\
A. Hebecker, Nucl. Phys. {\bf B632}, 101 (2002) [hep-ph/0112230]. 

\bibitem{Dudas} 
E. Cremmer, S. Ferrara, C. Kounnas, D.V. Nanopoulos, Phys. Lett. {\bf B133}, 61 (1983); 

\bibitem{String}
E. Witten, Phys. Lett. {\bf B155}, 151 (1985);\\ 
U. Ellwanger and M. G. Schmidt, Nucl. Phys. {\bf B294}, 445 (1987);\\ 
S. Ferrara, C. Kounnas, M. Porrati, Phys. Lett. {\bf B181}, 263 (1986);\\  
S. Ferrara, C. Kounnas, and F. Zwirner, Nucl. Phys. {\bf B429}, 589 (1994);\\ 
E. Dudas and C. Grojean, Nucl. Phys. {\bf B507}, 553 (1997).

\bibitem{Chang}
N. Chang, S. Ouvry, and X. Wu, Phys. Rev. Lett. {\bf 51}, 327 (1983).

\bibitem{Ellwanger}
U. Ellwanger, N. Dragon, and M. G. Schmidt, Phys. Lett. {\bf B145},
192 (1984); 
Nucl. Phys.{\bf B} {\bf 255}, 544 (1984); Prog. Part. Nucl. Phys. {\bf 18}, 1 (1987). 

\bibitem{Tytgat}
M. Faibarn, L. Lopez-Honorez, and M.H.G. Tytgat, hep-ph/0302160.

\bibitem{ourprep}
R. Hofmann, F. Paccetti Correia, M.G. Schmidt, Z. Tavartkiladze, In
preparation.

\bibitem{therm}
D.G.C. McKeon and A. Rebhan, Phys. Rev. {\bf D47}, 5487 (1993);\\ 
M. Haack and M.G. Schmidt, Eur. Phys. J. {\bf C7}, 149 (1999). 

\bibitem{outWL}
M. Mondragon, L. Nellen, M.G. Schmidt, and Ch. Schubert, 
Phys. Lett. {\bf B351}, 200 (1995) [hep-th/9502125];\\ 
{\sl ibid} {\bf B366}, 212 (1996) [hep-th/9510036].

\bibitem{Gersdorffgrav}
G.v. Gersdorff and M. Quiros, Phys. Rev. {\bf D65}, 064016 (2002) [hep-th/0110132];\\ 
G.v. Gersdorff, M. Quiros, and A. Riotto, Nucl. Phys. {\bf B634}, 90 (2002) [hep-th/0204041]. 

\bibitem{Kugo}
For a recent formulation of 5D SUGRA see: 
M. Zucker, Nucl. Phys. {\bf B570}, 267 (2000) [hep-th/9907082], 
Phys. Rev. {\bf D64}, 024024 (2001) [hep-th/0009083];\\ 
T. Kugo and K. Ohashi, Prog. Theor. Phys. {\bf 108}, 203 (2002) [hep-th/0203276].   

\end{thebibliography}

\end{document}